\documentclass{aastex631}
\usepackage{csquotes}

\begin{document}


\title{Unprecedented Multipoint Observation of Spatially Varying ICME Turbulence of Different Ages during October 2024 Extreme Solar Storm at 1 AU}

\author[0009-0005-7637-8346]{Shibotosh Biswas}
\affiliation{Space Physics Laboratory, Vikram Sarabhai Space Centre, Indian Space Research Organization,  Trivandrum,  Kerala 695022, India}
\affiliation{Department of Physics, University of Kerala, Trivandrum, Kerala 695581, India}
\correspondingauthor{biswas.sb655@gmail.com}

\author[0000-0003-4281-1744]{Ankush Bhaskar}
\affiliation{Space Physics Laboratory, Vikram Sarabhai Space Centre, Indian Space Research Organization, Trivandrum,  Kerala 695022, India}

\author[0009-0004-6143-8986]{S G Abitha}
\affiliation{Department of Physics, University of Kerala, Trivandrum, Kerala 695581,  India}

\author[0000-0002-4862-4141]{Omkar Dhamane}
\affiliation{National Institute for Astrophysics, Institute for Space Astrophysics and Planetology, Via del Fosso del Cavaliere 100, I-00133 Rome, Italy}

\author[0000-0002-6302-438X]{Sanchita Pal}
\affiliation{Centre for Space Science and Technology, Indian Institute of Technology Roorkee, Roorkee, 247667 Uttarakhand, India}

\author[0000-0003-2693-5325]{Dibyendu Chakrabarty}
\affiliation{Physical Research Laboratory, Ahmedabad-380009, India}

\author[0000-0002-1470-8443]{Vipin K Yadav}
\affiliation{Space Physics Laboratory, Vikram Sarabhai Space Centre, Indian Space Research Organization, Trivandrum, Kerala 695022, India}

\begin{abstract}
Understanding turbulence in interplanetary coronal mass ejections (ICMEs) is fundamental to space plasma research and critical for assessing the impact of space weather on geospace. Turbulence governs energy cascade, plasma heating, magnetic reconnection, and solar wind–magnetosphere coupling, thereby influencing both ICME evolution and geoeffectiveness. While previous event-based and statistical studies have examined ICME turbulence and its radial evolution in great detail, no significant measurements of ICME magnetic turbulence at a single vantage point have been obtained from multiple observatories separated azimuthally. Here, we present the first multipoint analysis of magnetohydrodynamic (MHD) turbulence across ICME plasma regions, using four spacecraft at the Sun-Earth L1 point, separated by $\approx80 R_{E}$ (meso-scale) along the dawn-dusk direction. Using high-resolution magnetic field observations from ISRO's Aditya-L1, NASA's Wind and ACE, and NOAA's DSCOVR, we analyze turbulence associated with the $10^{th}$ October 2024 solar storm, which triggered the second-strongest geomagnetic storm of solar cycle 25. Our results reveal significant variability and differing turbulence maturity across small separations,  supported by analysis of field-aligned and perpendicular magnetic-field cascades, indicating strong anisotropies. Sheath turbulence is substantially modified by shock-induced energy injection. Evidence of compressible turbulence and plasma energization at the flux rope interaction region indicates that internal processes, such as magnetic reconnection, strongly influence ICME plasma evolution, highlighting pronounced spatial variability in turbulence and plasma states observed by multiple L1 monitors near Earth and underscoring their potential role in space weather impacts.

\end{abstract}


\section{Introduction} \label{sec:intro}

Turbulence is a fundamental phenomenon in hydrodynamic and MHD systems that governs the transport, dissipation, and energy transfer across a wide range of spatial and temporal scales. It arises naturally and refers to the fluid's chaotic, irregular, multiscale motion. A defining feature of turbulence is the cascade of energy injected at large scales towards progressively smaller scales through non-linear interactions. In magnetized plasma, turbulence is more complex than in a neutral fluid due to the coupling between electromagnetic fields and particle dynamics. From laboratory plasmas to interplanetary \citep{bruno2013solar, Bruno_2016} and astrophysical plasmas \citep{Elmegreen_2004}, turbulence plays a central role in various processes, such as the heating and acceleration of charged particles \citep{Yordanova_2021, Chandran_2013}, momentum transfer, and mixing. Kolmogorov described the energy cascade in the inertial range (the scale between energy-injection and dissipation range) of neutral fluids, where neither external forcing nor viscous dissipation directly influences the dynamics \citep{frisch1995turbulence}. It was assumed that the statistical properties of turbulence are determined solely by the rate of energy dissipation per unit mass and display a power-law behavior for the energy spectrum as a function of wavenumber k: $E(k) \propto k^{-5/3}$, where $E(k)$ is spectral energy density. In 1965, Kraichnan \citep{kraichnan1965inertial} proposed a model of MHD turbulence, highlighting that energy transfer is governed by interactions between counterpropagating Alfvén waves along the magnetic field, which slow down the nonlinear cascade and extend the characteristic timescale from the eddy turnover time to the longer Alfvén crossing time, resulting in a slightly shallower energy spectrum. The expression for the power law in the Iroshnikov-Kraichnan \citep{kraichnan1965inertial, iroshnikov1964turbulence} theory is given by $E(k) \propto k^{-3/2}$. The high Reynolds number of the solar wind compared to that of ordinary fluids suggests that dissipation occurs through waves and instabilities rather than through viscosity. At the same time, the presence of a magnetic field introduces anisotropy, causing the turbulent eddy to manifest as Alfvén waves \citep{Alfven1942}. \\

The solar wind, a continuous flow of plasma from the Sun, is a collisionless and turbulent medium. It exhibits varying degrees of turbulence and distinct power-law trends at different scales \citep{bruno2013solar, telloni2016linking}. A detailed understanding of solar wind turbulence can be found in \cite{Bruno_2016}. In addition to the continuous solar wind, the Sun also ejects large-scale plasma with magnetic fields, known as coronal mass ejection (CME), from the corona. The interplanetary counterpart of CMEs (ICMEs, \cite{kilpua2017coronal}) constitutes the most consequential large-scale transients in the heliosphere and is the principal driver of heliospheric plasma variabilities and intense planetary geomagnetic disturbances \citep{Gopalswamy_2007}. ICMEs, often supersonic and super-Alfv\'enic, generate shock waves in front, characterized by abrupt enhancements in magnetic field, solar wind velocity, plasma density, and temperature \citep{gosling1973anomalously}. Between the shock and the ejected coronal mass, the solar wind is highly compressed, forming the ICME sheath, which exhibits rapid variations in the interplanetary magnetic field (IMF) and elevated plasma density and temperature. The ejected coronal mass is the large-scale, coherent magnetic flux rope, often called magnetic cloud(MC), exhibits enhanced magnetic field strength compared to the solar wind background \citep{burlaga1981magnetic}, rotation of magnetic field components \citep{klein1982interplanetary}, reduced proton temperature, and plasma beta (ratio between thermal and magnetic pressure) \citep{Nieves_Chinchilla_2018}. \\ 

Turbulent fluctuations in ICMEs are primarily driven by shocks, compression, and flux-rope structures at the largest energy-containing scales \citep{burlaga1991intermittent}.
In the sheath, perpendicular fluctuations are often amplified by shocks and plasma compression \citep{zank2002interaction}. In contrast, within MCs, the balance between parallel and perpendicular components is influenced by the flux rope's geometry. The presence of discontinuities and internal current sheets \citep{telloni2021evolution, Biswas_2025} often leads to enhanced dissipation in MC as well as ICME sheath \citep{Ghag_2026}. ICME magnetic turbulence plays a crucial role in determining its geoeffectiveness and impact on space weather. \cite{Thampi_2025} showed that the enhanced geoeffectiveness of the October 2024 geomagnetic storm occurred during the turbulent sheath phase. In the sheath region, young turbulence enhances magnetic variability and affects reconnection at Earth’s magnetopause, often intensifying geomagnetic storms \citep{kilpua2017coronal, Borovsky2020}. Inside MCs, turbulence gradually ages, leading to steeper spectral slopes and reduced intensity, yet it still influences flux-rope stability and particle transport (e.g., \cite{bruno2013solar, chen2020effects}). These evolving turbulence regimes regulate both the onset and duration of geomagnetic disturbances, thereby affecting forecasting accuracy \citep{Zank2016, telloni2021evolution}. By incorporating turbulence characteristics into space weather models, it becomes possible to enhance our estimation of the timing, strength, and persistence of ICME-driven events that affect technological systems around and on Earth \citep{Lugaz2017, Manchester2017}.\\

Myriad studies have been conducted to date on MHD and kinetic-scale turbulence in ICMEs \citep{Shaikh_2024, M_rquez_Rodr_guez_2023, Sorriso_Valvo_2021, Good_2023, Dhamane_2024, dhamane2023observation, Raghav_2023}. In this era of space exploration, we use in situ measurements from multiple observatories at different vantage points in the interplanetary medium between the Sun and Earth. Such opportunities open new avenues for investigating fundamental physical processes in great detail. The radial alignment of spacecraft provides a unique opportunity to study the evolution of turbulent plasma and its dynamic behavior as it propagates towards Earth \citep{Good_2020}. However, ICME magnetic turbulence studies from a single vantage point, using multiple observatories aligned azimuthally, are not investigated in detail. The magnetic clouds, often modeled as twisted flux ropes, exhibit coherence in a small angular separation \citep{Owens_2020, Owens_2017}, although they can show significant variation in the case of complex interacting ICMEs \citep{Pal_2025}. Therefore, the turbulence characteristics may vary significantly in a small separation, also providing information about the evolution of the ICME plasma in a distinctly different way. Using in-situ observations from ISRO, NASA, and NOAA's Sun-Earth L1 monitors, we analyze the spatio-temporal characteristics of turbulence across distinct plasma regions within the ICME. The paper proceeds by describing the data and methodology in Section \ref{sec:data}, followed by the results and discussion in Section \ref{sec:results}, and concluding in Section \ref{sec:conclusion}. \\ 


\section{Data and Methodology}\label{sec:data}

Aditya-L1 is India's first dedicated solar mission \citep{TripathiAdityaL1SUIT, ISROAdityaL1}, launched by the Indian Space Research Organisation (ISRO). Data were obtained from the fluxgate magnetometer (MAG) \citep{Yadav2025} with a time resolution of 1 second, which has been used in a previous study of \cite{Yadav_2025}. The plasma moments and differential flux of protons have been obtained from the solar wind ion spectrometer of Aditya Solar Wind Particle Experiment (ASPEX-SWIS, \cite{Goyal_2018, Kumar_2025}) instrument onboard Aditya-L1. From NASA's Advanced Composition Explorer (ACE) \citep{garrard1998ace}, level 2 data of magnetic field  (1-second cadence, \cite{https://doi.org/10.48322/7xyh-4z44}) and plasma parameters (5-minute cadence, \cite{https://doi.org/10.48322/pfr6-fg57}) have been obtained from MAG and SWEPAM (Solar Wind Electron, Proton, and Alpha Monitor) instruments. . The Deep Space Climate Observatory (DSCOVR) is a joint mission of NASA and NOAA \citep{DSCOVRMissionOverview}. We used magnetometer data with a 1-second time resolution. Moreover, high-resolution 92 ms data from the magnetic field investigation (MFI, \cite{https://doi.org/10.48322/0v0h-df27}) and the three-dimensional plasma and energetic particle investigation (3DP) \citep{https://doi.org/10.48322/kwfz-zk29} onboard Wind spacecraft were used. All data from ACE, Wind, and DSCOVR are publicly available at \url{https://cdaweb.gsfc.nasa.gov/}, and the data DOIs are provided for the corresponding instruments mentioned above. The Aditya-L1 mission data are available at \url{https://www.issdc.gov.in/adityal1.html}.\\

Analysis of the power spectrum of magnetic field fluctuations ($\delta B = B - <B>$, where $<B>$ is the running average of the magnetic field) is an invaluable tool for examining the time–frequency evolution of turbulence. The Power spectral density (PSD) quantifies how the variance (or energy) of a fluctuating signal is distributed across different frequencies \citep{bendat1986decomposition}. PSD analysis provides an averaged spectrum over a given interval, showing how fluctuation power varies across frequencies as the spacecraft moves through the ICME sheath, MC, and background solar wind. We used the highest-resolution magnetic field data available (1-s cadence, except for Wind's 92-ms data) from the L1-orbiting spacecraft and calculated the magnetic field turbulence component ($\delta B$). The PSD of all the field components can be expressed as $P_{x} = |\delta B_{x}(f)|^2$, $P_{y} = |\delta B_{y}(f)|^2$, and $P_{z} = |\delta B_{z}(f)|^2$, where $\delta B_{x}(f)$, $\delta B_{y}(f)$, $\delta B_{z}(f)$ represents the fourier-ternsformed components. The PSD of the trace of the magnetic field is calculated by $P_{tr} = P_{x}+P_{y}+P_{z}$. We used the Welch method \citep{Welch_1967} to compute the PSDs.   \\
Turbulence characteristics in space plasmas are shaped by the large-scale magnetic field, which provides a preferred natural direction for fluctuations to develop \citep{matthaeus1990evidence}. To study this directional dependence, the field-aligned coordinate (FAC) system is used, separating IMF fluctuations into parallel and perpendicular components relative to the local mean magnetic field (e.g., \cite{horbury2008anisotropic}). We utilize a 5-minute centered rolling window to estimate the local mean magnetic field. At a 1 Hz (11 Hz for Wind) sampling rate, this corresponds to an exact window size of 301 (3301 for Wind) data points. Subtracting this 5-minute moving average acts as an effective high-pass filter with a cutoff frequency well below the chosen primary inertial range for spectral fitting (0.01 to 0.1 Hz). It ensures that the exact turbulent fluctuations ($\delta B$) accurately reflect the true MHD fluid cascade, without being suppressed by background-field subtraction. The parallel and perpendicular magnetic field components are then used to calculate the power spectral density, as described in the preceding paragraph. The power spectral density varies with frequency as a power law. To analyze the nature of the turbulence cascade, the slope ($\alpha$) of the spectral density variation is estimated using power-law fitting. This approach is particularly valuable in ICME studies, because turbulence cascades show noticeable directional preference (strong perpendicular in case of ideal incompressible MHD turbulence) which is established by anisotropic turbulence theories \citep{goldreich1995toward, chen2016anisotropic}. In contrast, parallel fluctuations are more closely linked to compressive modes and wave–particle interactions, which are essential for kinetic-scale energy dissipation \citep{sahraoui2010three}.

\section{Results and Discussion} \label{sec:results}

\subsection{Overview of the event}

\begin{figure*}[h]
\centering
\includegraphics[width=\linewidth]{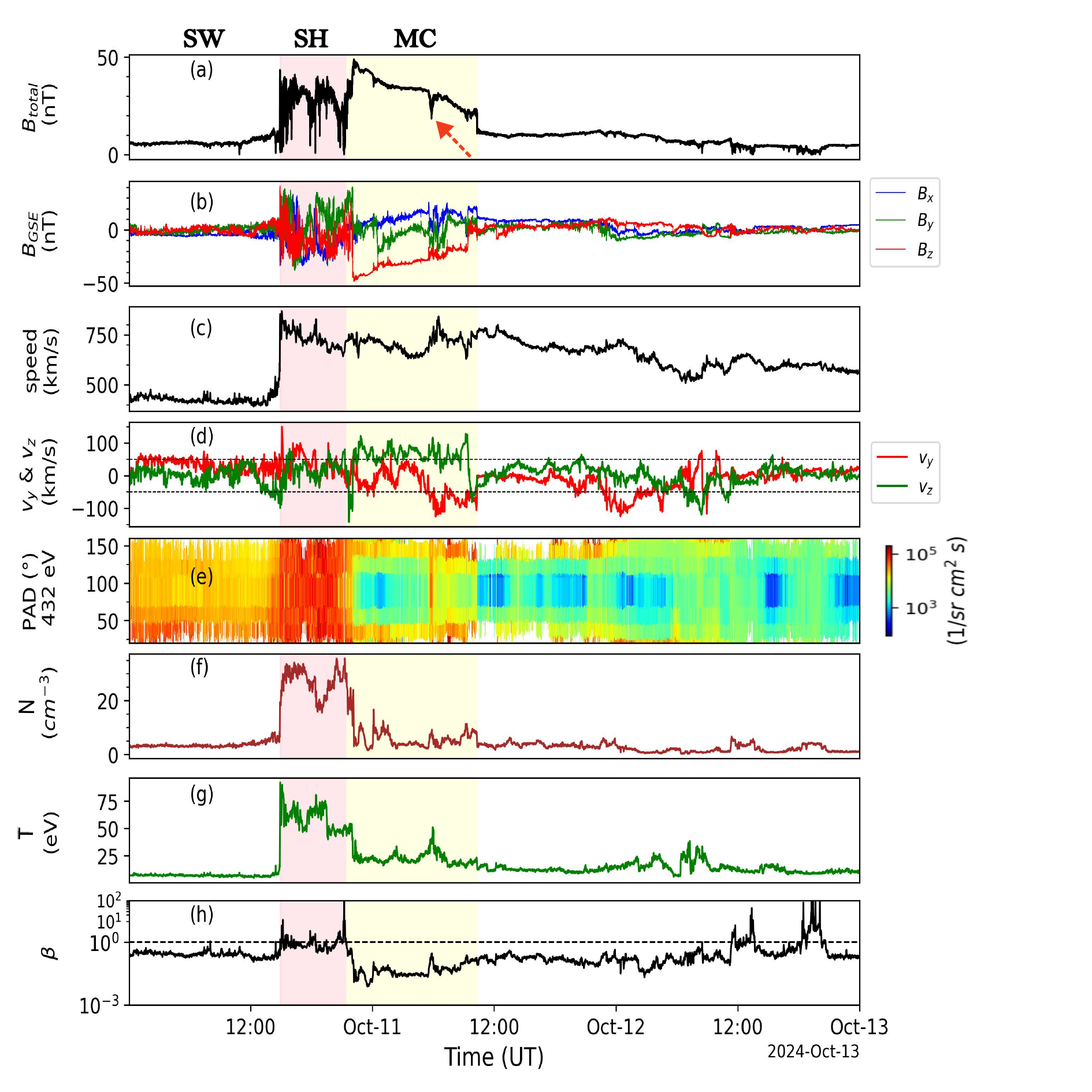}
\caption{Solar wind and ICME plasma and magnetic field parameters measured by the Wind spacecraft during the extreme solar storm from 10 to 13 October 2024. Panels a-h represent magnetic field magnitude, components in geocentric solar ecliptic coordinates, solar wind bulk speed, proton velocity components along $GSE_{y}$ and $GSE_{z}$, suprathermal pitch angle distribution(432 eV channel), proton density,  temperature, and plasma $\beta$. The Solar wind (SW), ICME sheath (SH), and magnetic cloud (MC) regions are marked above with their corresponding abbreviations. The sheath and magnetic cloud are highlighted in light pink and yellow, respectively. The dip in the magnetic field within MC at 06:00 UT, October 11, is indicated using a red arrow in panel a.}
\label{fig:Wind summary}
\end{figure*}

\begin{figure}[h]
    \centering
    \includegraphics[width=0.9\linewidth]{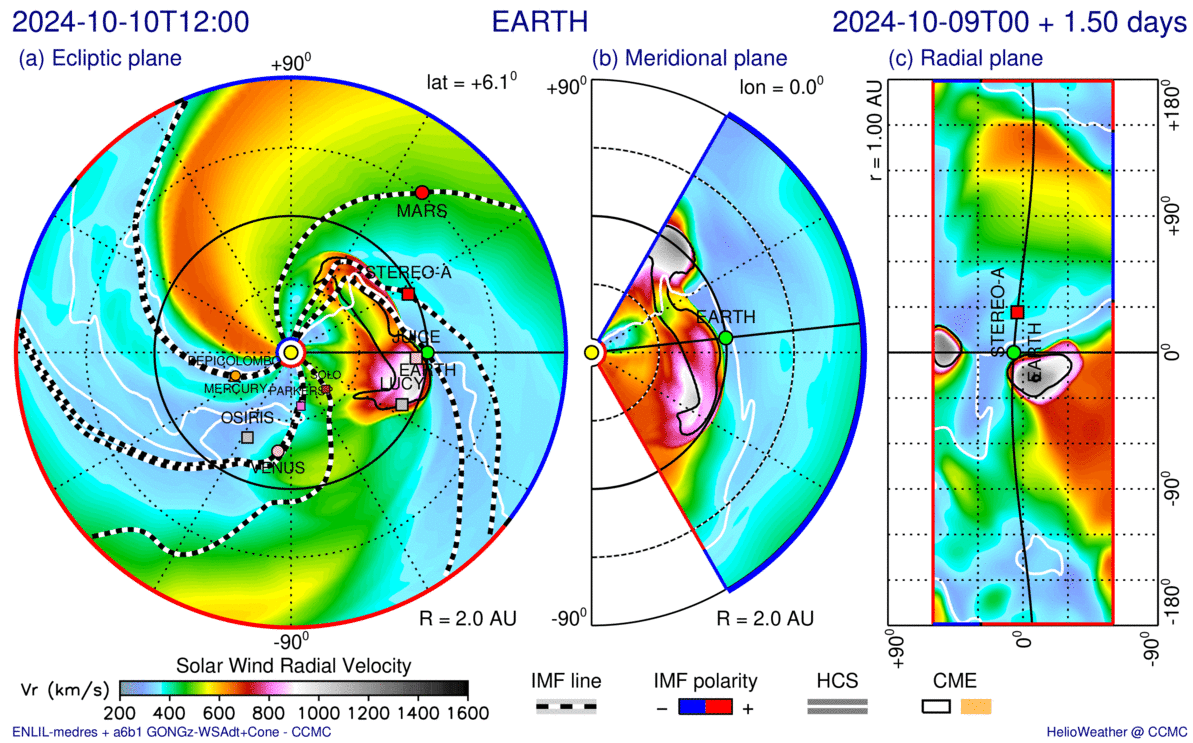}
    \caption{WSA-ENLIL simulated snapshot of CME in the heliosphere at arrival time at Earth. The solar wind radial velocity is shown in the panels. The complete simulations can be accessed through the URL: \url{https://ccmc.gsfc.nasa.gov/ror/results/viewrun.php?runnumber=Ankush_Bhaskar_122525_SH_1 }}
    \label{fig:enlil}
\end{figure}

\begin{figure*}[htbp]
\centering
\includegraphics[width=0.8\linewidth,height=8cm]{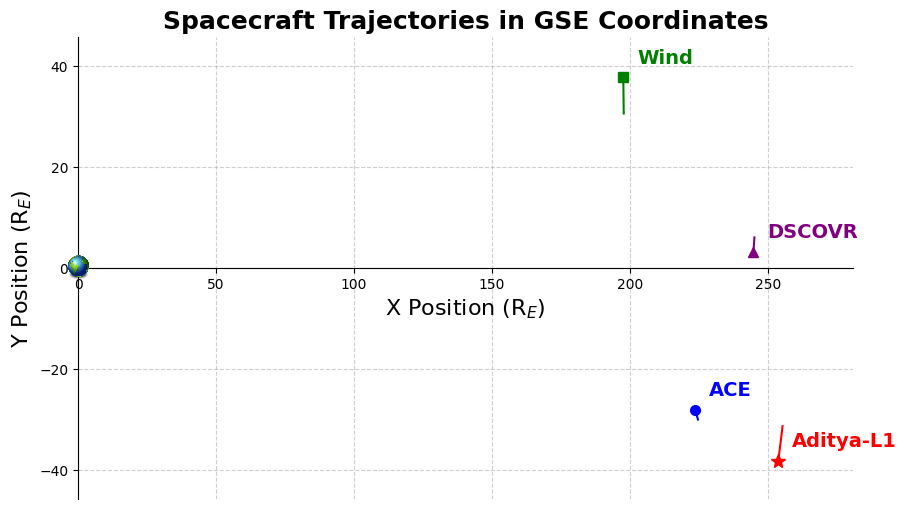}
\caption{Position of Aditya-L1, ACE, DSCOVR, and Wind in GSE (Geocentric Solar Ecliptic) coordinates during the ICME event from October $10^{th}$ to $12^{th}$, 2024}
\label{fig:spacecraft loc}
\end{figure*}

From October 10-14, 2024, solar coronal mass ejections affected Earth's magnetosphere, resulting in the second-strongest geomagnetic storm of solar cycle 25. Figure~\ref {fig:Wind summary} demonstrates the high-resolution IMF and plasma parameters obtained from the Wind spacecraft. The panels display the IMF magnitude and components, solar wind velocity components in the geocentric solar ecliptic (GSE) coordinate system, proton density, and temperature, respectively, from 10-13 October 2024. The sudden changes in IMF, bulk velocity, density, and temperature indicate the shock arrival after 14:00 UT, followed by a turbulent region characterized by fluctuating IMF magnitude and components, elevated proton density, and temperature. After 10th October at 22:00 UT, IMF-$B_{z}$ underwent a sudden transition to negative, followed by a smooth variation of other magnetic field components. The reduced proton density and temperature, as well as plasma $\beta$, suggest the presence of an ICME MC. Between 11 October at 06:00 UT to 12 October at 12:00 UT, we observe increments of $v_{y}$ and $v_{z}$ above -50 km/s (figure~\ref{fig:Wind summary}d) along with sustained field-aligned suprathermal electron pitch-angle distribution (figure~\ref{fig:Wind summary}e), and low ion $\beta$. These might indicate multiple ICMEs, although the evidence is inconclusive and further investigation is necessary. A noticeable dip in IMF magnitude, abrupt magnetic field variation, and local proton density enhancement were observed on 11 October at 06:00 UT within the highlighted MC, which will be further discussed in a separate section.  \\
The simulated CME propagation using the physics-based Wang-Sheeley-Arge (WSA)-Enlil model is shown in Figure \ref{fig:enlil}. The CME propagation is simulated based on initial CME parameters obtained from DONKI (\url{https://kauai.ccmc.gsfc.nasa.gov/DONKI/search/}). The model's prediction of the CME's arrival at Earth is nearly consistent with the observations. Further details and simulation outputs could be retrieved from \url{https://ccmc.gsfc.nasa.gov/ror/results/viewrun.php?runnumber=Ankush_Bhaskar_122525_SH_1}. Several spacecraft in the interplanetary medium and near Earth recorded this event.
Separation of the L1 spacecrafts along the GSE y-axis enables us to investigate the spatial variation of turbulence in detail using a multi-point observational approach. Figure~\ref{fig:spacecraft loc} describes the spacecraft locations with the trajectory from 10th to 12th October in the GSE x-y plane. Aditya L1, ACE, DSCOVR, and Wind are systematically separated in the Y-direction from dawn to dusk. Two observatories, specifically Wind and Aditya-L1, are most separated by approximately 80 $R_{E}$ in the Y-direction.

\begin{figure*}[h]
\centering
\includegraphics[width=0.9\linewidth]{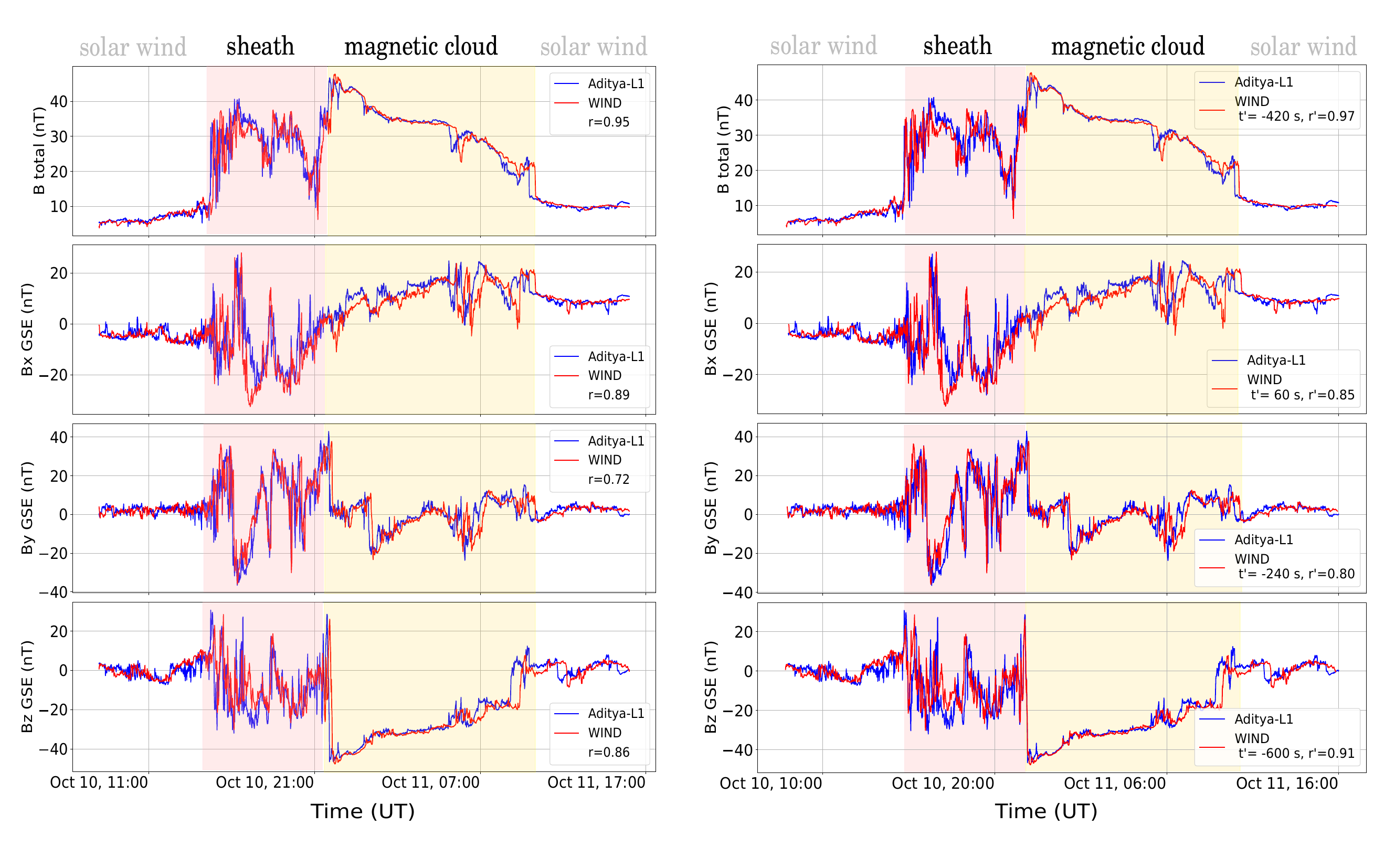}
\caption{This figure shows the correlation of total magnetic field and respective components in the GSE coordinate of Wind and Aditya-L1, having the highest separation between them. The left panel displays the uncorrected correlation, whereas the right panel displays the time-lag-corrected correlation.}
\label{fig:correlation}
\end{figure*}

We performed a correlation analysis of magnetic field components between Aditya-L1 and Wind to assess local-scale variability on the order of tens of $R_{E}$. Figure~\ref{fig:correlation}(a) shows magnetic field intensity and components of Aditya-L1 and Wind overplotted on top of each other. The solar wind, sheath, and MC are identified using magnetic field observations. The three distinct regions highlighted in white, pink, and yellow correspond to low-intensity background IMF (solar wind), high-fluctuation (sheath), and smooth variation (MC), respectively. Aditya-L1 first observed the passage of ICME owing to its position ahead (towards +x GSE) of Wind, noticeable in Figure~\ref{fig:correlation}a. The estimated cross-correlation coefficient is statistically significant (p=0.00), and the initial analysis (without time adjustment) shows a high correlation. After accounting for the time lag between the spacecraft, calculated based on the time at which the correlation between the total magnetic field of Aditya-L1 and Wind is maximum (Figure~\ref{fig:correlation}), the IMF magnitude shows a robust match, with a correlation coefficient (r) of 0.99. The spacecraft separation is significant along the GSE y direction, so even small spatial gradients or inhomogeneities may lead to variation in the $B_y$ component. Moreover, anisotropies in magnetic-field turbulence also lead to weaker correlations, indicating that the east-west component is more sensitive to small-scale fluctuations, turbulence, and wave activity. The north-south component ($B_z$ GSE) shows substantial improvement after lag correlation, which is responsible for the critical geomagnetic activity and space weather effects. The radial component ($B_x$ GSE) also remains highly correlated, though with a slight reduction after adjustment. The correlation analysis results are displayed in the table~\ref{tab:correlation}. Overall, both spacecraft have observed similar large-scale structures, but there are notable differences in the small-scale magnetic field fluctuations, which are the primary interest of the current study.

\begin{deluxetable}{lccc}
\tablecaption{Correlation analysis of magnetic field components observed by Aditya-L1 and Wind. The table lists the initial correlation coefficient $r$, time lag, and the adjusted correlation coefficient $r'$ after applying the time lag.\label{tab:correlation}}
\tablehead{
\colhead{Variable} & 
\colhead{\begin{tabular}{c} Initial \\ Correlation \\ coefficient $r$ \end{tabular}} & 
\colhead{\begin{tabular}{c} Time Lag \\ (s) \end{tabular}} & 
\colhead{\begin{tabular}{c} Time-Lag Adjusted \\ Correlation \\ coefficient $r'$ \end{tabular}}
}
\startdata
$B_\mathrm{total}$ & 0.95 & $-420$ & 0.99 \\
$B_x$ (GSE)        & 0.89 & $+60$  & 0.86 \\
$B_y$ (GSE)        & 0.72 & $-240$ & 0.81 \\
$B_z$ (GSE)        & 0.86 & $-600$ & 0.94 \\
\enddata
\end{deluxetable}

\subsection{Magnetic Energy Spectrogram}

\begin{figure*}[h]
\centering
\includegraphics[width=0.9\linewidth]{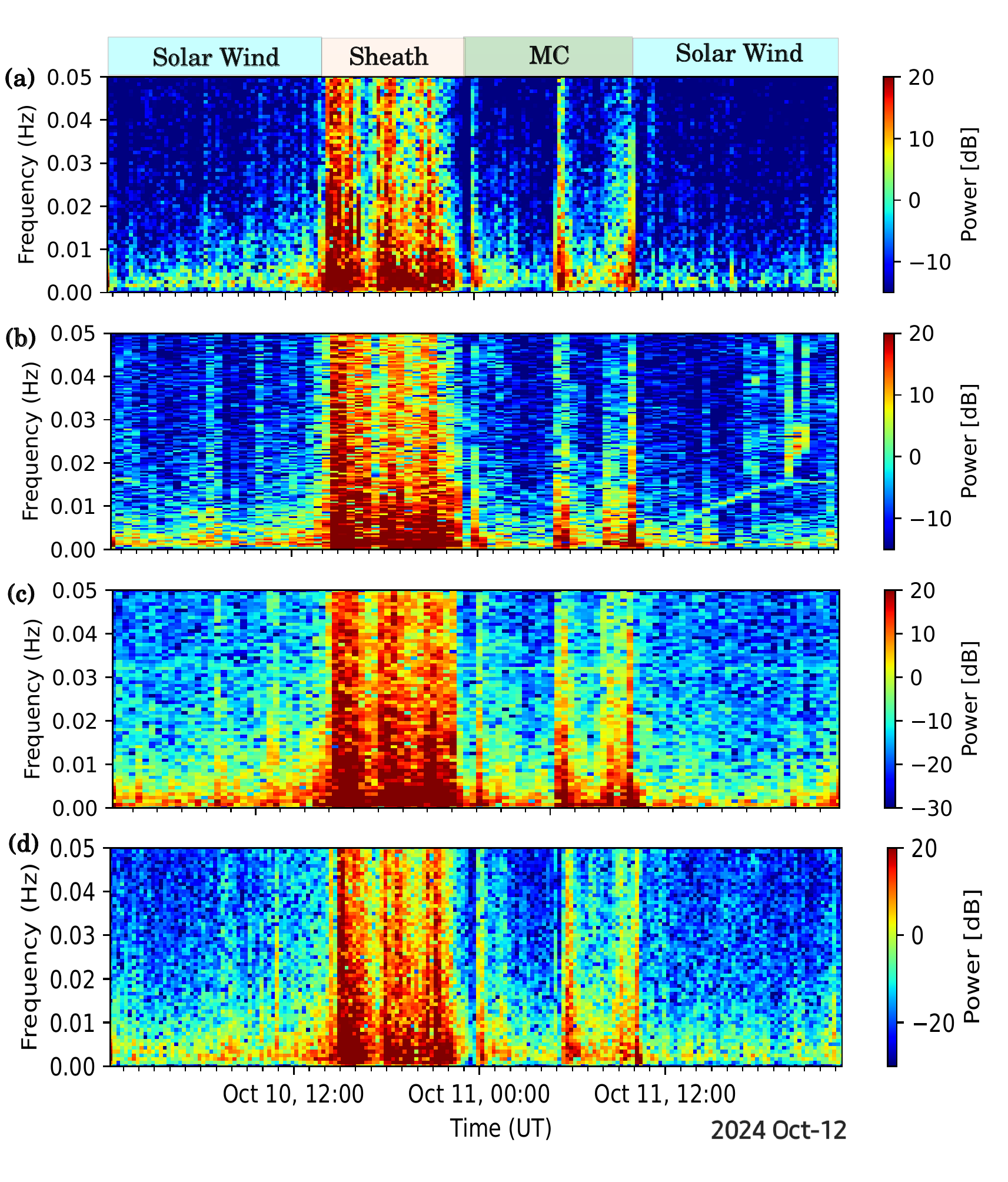}
\caption{The magnetic power spectra of the four spacecraft are displayed in this figure, with panels a to d representing DSCOVR, Aditya-L1, ACE, and Wind, respectively. All magnetic field data have been resampled at 10-second time resolution. The solar wind, sheath, and magnetic cloud can be distinctly identified with their respective turbulence power characteristics. }
\label{fig:powerspectrum}
\end{figure*}

The spectrogram analysis of magnetic field magnitude from multiple spacecraft at the Lagrangian point 1 revealed key characteristics of turbulence. Figure~\ref{fig:powerspectrum} demonstrates that different ICME substructures exhibit unique spectral signatures as the system passes through the observatories in sequence. The background solar wind shows fluctuating power concentrated at lower frequencies, suggesting fluid-scale (MHD) turbulence (Figure~\ref{fig:powerspectrum}). It is noted that the magnetic power spectrum for all spacecraft has been computed from resampled magnetic field data with a 10s cadence, which constrains the maximum frequency to 0.05 Hz. The ICME sheath is the compressed solar wind resulting from the supersonic flow of the ejected CME. Due to the continuous presence of drivers (e.g., shock compression, discontinuities, and current sheets), the sheath region is highly turbulent. The fluctuation power is highest and distributed across a broad frequency range, indicating intense turbulent activity driven by shock compression and plasma irregularities. This was marked by sharp spikes in power at the shock front and sustained enhanced power throughout the sheath. However, the magnetic cloud exhibits reduced power and a narrower frequency range, reflecting its more coherent, organized magnetic structure and suppressed turbulence. Moreover, we occasionally observe an increase in the power spectrum across a broad frequency range within the magnetic cloud, one of which we discuss separately in section~\ref{sec:results} (on October 11, 06:00 UT). The remaining two intense bands of high turbulence power at the sheath-MC boundary (00 UT, october 11) and the tail-end of marked MC (10:00 UT, october 11) are possibly owing to the sheath-MC boundary interaction, where we observe a sharp negative jump in IMF $B_z$, $B_y$ and  (figure~\ref{fig:Wind summary}b)  and tail compression (on 11th October at 10:00 UT). Power spectrum analysis provides broad information for identifying turbulent regions in space plasma, but it limits understanding of the specific stages of cascading involved.

\subsection{Power Spectral Density Analysis}

\begin{figure*}[h]
    \centering
    \includegraphics[width=\textwidth]{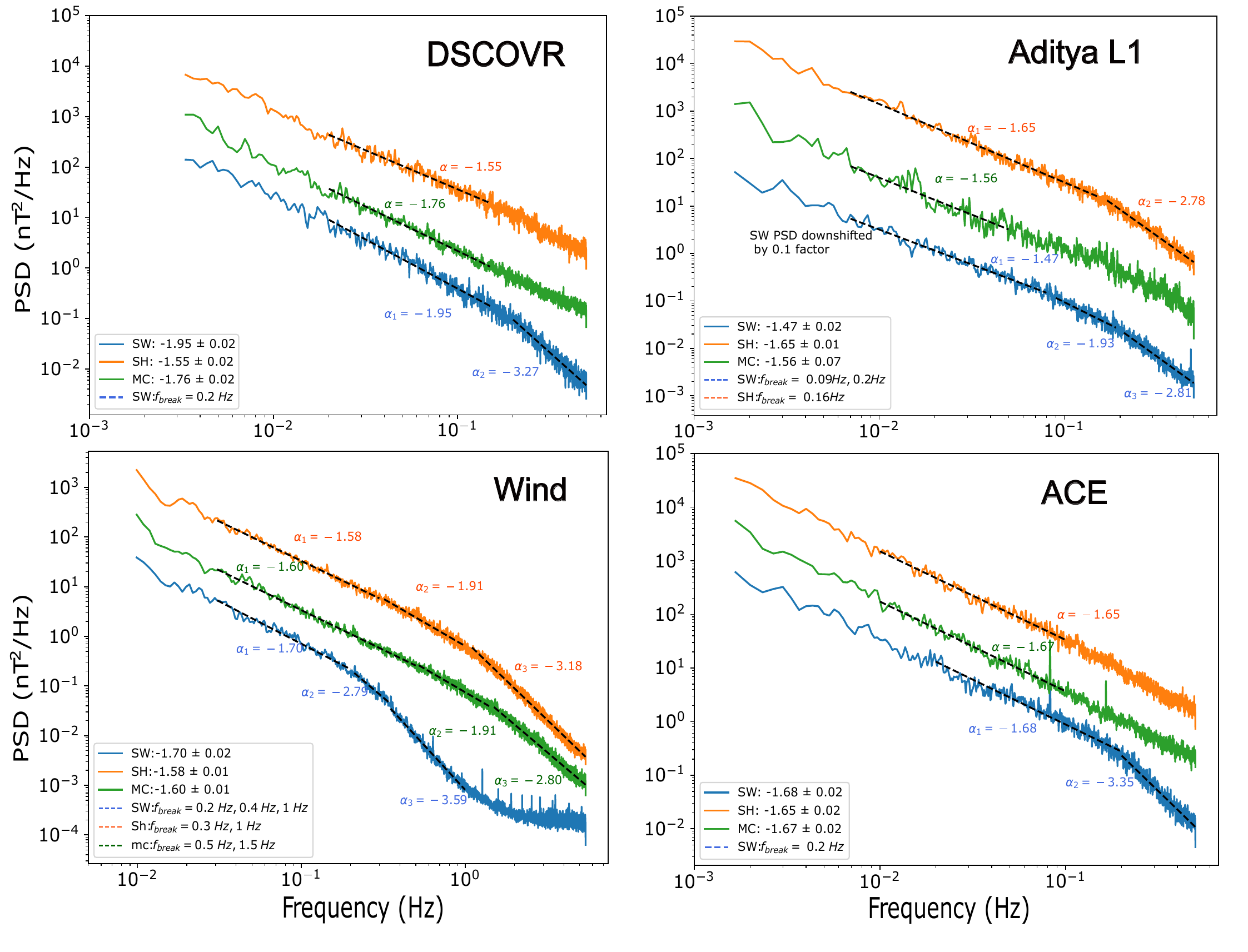}
    \caption{Power spectral density (PSD) slopes across different solar wind regions: ambient solar wind (blue), sheath (orange), and magnetic cloud (green). For all spacecraft, PSD is computed from 1 Hz data, except for Wind, which is computed from 11 Hz data. The slopes have been fitted manually in different frequency scales. For Wind and Aditya-L1, an additional frequency range is observed, with a different spectral index. }
    \label{fig:allregionpsd}
\end{figure*}

Figure~\ref{fig:allregionpsd} shows multipoint analysis of the PSD of the trace of the magnetic field with their respective spectral slopes. We selected a fixed frequency range of 0.01-0.1 Hz (for all three regions) for primary inertial scaling ($\alpha_{1}$) across all three regions to ensure our fits capture pure macroscopic fluid-scale turbuence. The upper limit of 0.1 Hz was chosen because our estimates of ion kinetic scale frequencies in the spacecraft frame for the solar wind, sheath, and magnetic cloud consistently exceed 0.1 Hz. Therefore, it may serve as a safe upper boundary to avoid spectral steepening associated changes. Conversely, the lower limit of 0.01 Hz was established to ensure the analysis remains safely above the 1/f energy-injection break, which enables the power law fits to be applied strictly to the fully developed turbulence cascade \citep{bruno2013solar, Zhao_2021}, while also accommodating the finite duration of the ICME sub-intervals analyzed to ensure statistical stationarity and adequate spectral averaging \citep{Kilpua_2020}. While the differing data sampling rates (1 Hz and 11 Hz) in our datasets technically permit different fitting ranges, we restrict our analysis to this specific range to ensure a standardized, unbiased comparison across all instruments and regions. The upper limit of the upstream solar wind interval is set to 10 October 06:00 UT, with a time separation of about 2 hours to neglect foreshock effects. A noticeable difference in spectral slopes in the solar wind is evident across spacecraft in this figure. In table~\ref{tab:fac_analysis}, the spectral slopes are listed for all speacecrafts. Although the solar wind spectral slope in fluid-scale follows the Iroshnikov-Kraichnan (IK, $E(k) \propto k^{-3/2}$) and Kolmogorov ($E(k) \propto k^{-5/3}$) power law scaling at ACE and Aditya-L1 (PSD downshifted by a factor of 0.1 to avoid overlapping, figure~\ref{fig:allregionpsd}), respectively, it differs largely as measured by the other two spacecrafts. At Wind and DSCOVR, spectral slopes get significantly steeper in the inertial scale ($\approx -1.8$ on average). However, we observe a common spectral break at 0.2 Hz at all the solar wind observations separated by  $\approx 80~R_{E}$. 
The observed spectral break near 0.2 Hz at 1 AU solar wind data marks the fundamental transition from MHD inertial range to the ion kinetic regime. Because the turbulent cascade is highly anisotropic, this break frequency typically corresponds to a Doppler-shifted spatial boundary, such as the ion gyro radius or the inertial length. However, the subsequent steepening of the slope beyond this break is governed by a complex interplay between spatial structures and resonant wave-particle interactions \citep{Chen_2020, Telloni_2015, Wang_2018}. Our observations suggest the dominant dissipation mechanism depends strongly on the local environment. In the highly compressed sheath, the spectrum steepens dramatically ($\alpha \approx-3.18$), characteristic of dissipation by small-scale coherent structures and current sheets. Conversely, within the stable magnetic cloud, the suppression of such structures results in a shallower profile ($\alpha \approx-2.0$), indicating the progressive, resonant wave-particle interactions likely serve as the primary mechanism for energy transfer at sub-ion scales. It is worth noting that our interpretations are restricted to fluid-scale turbulence. In the sheath region, the spectral density is two orders of magnitude higher than solar wind, which is expected from figure~\ref{fig:powerspectrum}. The spectral slopes within the MHD scale exhibit a classical IK power-law behavior, except at Aditya-L1 and ACE, which are close to the Kolmogorov scaling, indicating a steady-state inertial-scale energy cascade. Such observations suggest continuous energy injection into the sheath region across all the spacecraft. Additionally, Aditya-L1 and Wind, two widely separated spacecraft, show spectral breaks in the sheath PSD at 0.1 and 0.3 Hz, respectively, indicating that local-scale plasma processes influence the turbulent energy cascade, thereby steepening the spectral slope and marking the transition to ion-kinetic scales as discussed before.

In contrast, the MC exhibits a single spectral slope close to the Kolmogorov scale, except for Aditya-L1, where the PSD variation is very irregular and the spectral power increases above 0.1 Hz (after the break point). This might imply energy deposition by specific plasma wave modes. Previous statistical studies have established that the turbulence inside MC frequently follows $f^{-5/3}$ spectral index at the inertial range, and the origin of the well-developed turbulence is the non-linear interaction of counterpropagating Alfvenic fluctuations \citep{Shaikh_2024, Good_2020}. However, local-scale plasma processes within MC and interactions among multiple flux ropes can significantly influence MC's turbulent evolution locally and globally \citep{Telloni_2020}. In the upcoming sections, we investigate a discontinuity in the MC and, in greater detail, how turbulence properties differ across ICME.

\begin{deluxetable*}{lccccc}
\tablecaption{Spectral slopes with standard error for solar wind, sheath, and magnetic cloud.\label{tab:psd_slopes}}
\tablehead{
\colhead{Satellite} & 
\colhead{Solar Wind} & 
\colhead{Sheath} & 
\colhead{Magnetic Cloud} & 
}
\startdata
Wind       & $-1.70 \pm 0.02$ & $-1.58 \pm 0.01$ & $-1.60 \pm 0.01$  \\
DSCOVR     & $-1.95 \pm 0.02$ & $-1.55 \pm 0.02$ & $-1.76 \pm 0.02$  \\
ACE        & $-1.68 \pm 0.02$ & $-1.65 \pm 0.02$ & $-1.67 \pm 0.02$  \\
Aditya-L1  & $-1.47 \pm 0.02$ & $-1.65 \pm 0.01$ & $-1.56 \pm 0.07$  
\enddata
\end{deluxetable*}

\subsection{Turbulence Power Anisotropy}

\begin{figure*}[h]
\centering
\includegraphics[width=\textwidth]{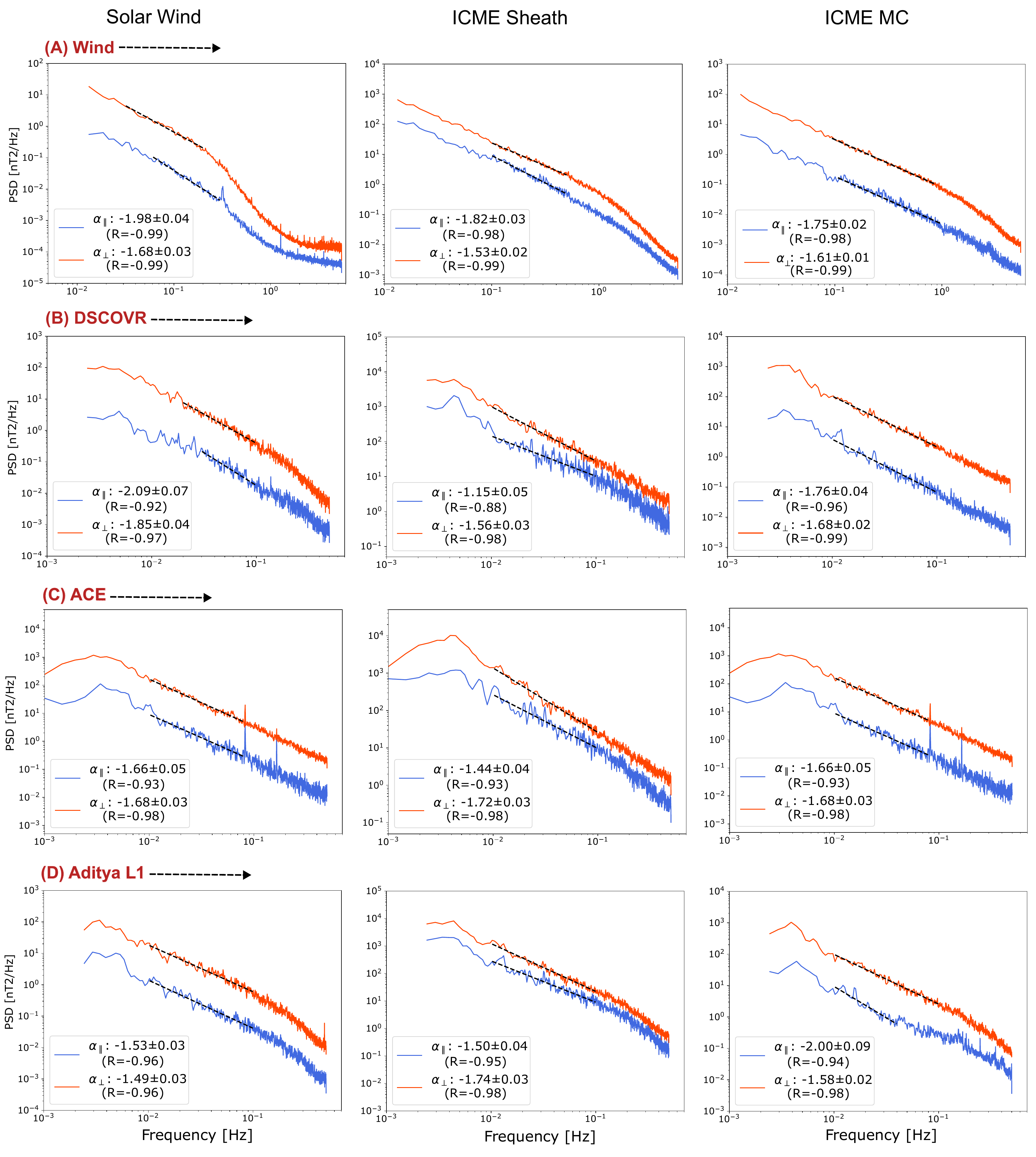}
\caption{Turbulence spectral plots of field-aligned and perpendicular magnetic fluctuations ($\delta B=B - <B>$, shown in blue and red color respectively) for the four spacecraft: Wind, DSCOVR, ACE, and Aditya-L1. The spectral slopes are shown in each panel by $\alpha_{\parallel}$, and $\alpha_{\perp}$. The Pearson correlation coefficient (R) for the power-law estimation is shown in each panel. The frequency range for the power-law fit has been chosen manually and is distinct for each case on the MHD scale. }
\label{fig:anisotropy}
\end{figure*}

The presence of a magnetic field significantly alters the properties of turbulence in a fluid, providing a natural direction for the efficient energy cascade and leading to anisotropy. Analysis of magnetic fluctuations resolved in field-aligned and perpendicular directions with respect to the background magnetic field reveals apparent differences in turbulence anisotropy across the spacecraft. \cite{goldreichsridhar1995} proposed an anisotropic turbulence model (GS95) accounting for the presence of a strong background magnetic field, which breaks spatial symmetry. The main assumption of GS95 is critical balance, which depict that the linear Alfv\'enic timescale ($\tau_{A} = l_{\parallel}/V_{A}$) is compareable to the non-linear eddy turnover time ($\tau_{nl} = l_{\perp}/v_{l}$), which lead to steeper parallel spectral slope (where $l_{\parallel}$ and $l_{\perp}$ are the characteristics dimension of turbulence eddy measured parallel (along), and perpendicular (acrosss) the local  agnetic field, and $v_{l}$ represents the velocity fluctuation amplitude of the eddy at perpendicular scale.)
However, GS95 theory deals entirely with the incompressible fluctuations ($\delta B_{\perp}$) and shows how the energy of Alfv\'enic fluctuations cascades along the magnetic field rather than across it. Here in this article, we compare the spectral slopes of $\delta B_{\parallel}$ (compressible fluctuations) and $\delta B_{\perp}$ to investigate how compressible turbulence evolves compared to the incompressible counterpart. In figure~\ref{fig:anisotropy}, we show parallel and perpendicular magnetic power spectral density plots for all four spacecraft. We introduce the component anisotropic scaling ratio, defined as $r_{A} = \alpha_{\perp}/\alpha_{\parallel}$, which helps us assess component equilibration. The ratio $r_{A}\approx1$ suggests the compressible and incompressible cascades are fully coupled, and energy is cascading through both modes at the exact same rates, $r_{A}>1$ represents compressible reset i.e. the compressible spectrum is shallower than the incompressible one, and finally $r_{A}<1$ suggests compressible steepening where compressible energy is either cascading down to smaller scales much faster than the Alv\'enic energy or it is experiencing stronger dissipation. For the full 3D anisotropic solar wind turbulence, refer to  \cite{Chen_2012}. \\

In the ambient solar wind, the scaling of compressible versus incompressible components varies significantly with the spacecraft's azimuthal separation, highlighting the solar wind's heterogeneous nature. The Wind and DSCOVR observed $r_{A}=0.85$ and $r_{A}=0.89$, respectively, suggesting high component anisotropy as the compressible turbulence is significantly steeper than the incompressible one. Moreover, a steeper perpendicular slope ($\alpha_{\perp}\approx-1.85$) measured by DSCOVR may indicate a higher degree of intermittency, which further steepens the cascade beyond the Kolmogorov steady-state rate, suggesting enhanced compressible dissipation in this stream of solar wind. On the other hand, ACE observes a similar energy cascade rate, with both parallel and perpendicular fluctuations following the Kolmogorov -5/3 scaling, which indicates steady-state turbulence. Finally, Aditya-L1 shows an Alfvénic turbulence cascade ($r_{A}=0.97$), characterized by the IK spectrum. This indicates a moderate turbulence maturity and suggests an equilibrated but less developed cascade.\\

The ICME sheath is the most complex environment in which the solar wind is compressed and heated by the ICME shock, and it can significantly affect the turbulence cascade. The sheath scaling anisotropy, as depicted in the Table~\ref{tab:fac_analysis}, shows a significant departure from the steady-state and reflects enhanced intermittency and compression.
Most spacecraft (DSCOVR, ACE, and Aditya-L1) exhibit a parallel slope extremely shallow, ranging from -1.15 to -1.50. This is a signature of young turbulence with fresh injection of energy. According to \cite{Zhao_2021}, the shock acts as a fresh energy source, creating a $ 1/f$-type variation that describes the energy-injection range. The injection of shock energy drives compressible fluctuations (creating large-scale variations in magnetic pressure) and resets the $\delta B_{\parallel}$ that hasn't had time to cascade into a steeper Kolmogorov steady state. The slope remains shallow because the energy reservoir at large scales remains dominant. In the slab and 2D model of nearly incompressible MHD (NIMHD) turbulence \citep{Adhikari_2015, Zank_2017, Adhikari_2022}, the perpendicular (2D) cascade and parallel (slab) cascade are processed differently. The perpendicular slopes are consistently steeper ($\alpha_{\perp} \approx-1.53~to~-1.74$) than parallel. This suggests that 2D eddies are compressed by shocks, which accelerate their nonlinear interactions, and that they quickly reach a Kolmogorov-like equilibrium state. The observed anisotropy inversion ($r_{A}>1$, except Wind) is a direct consequence of the shock selectively developing the perpendicular fluctuation and the parallel counterpart becoming younger. Even though all spacecraft are separated by only a few tens of $R_{E}$, Wind shows a remarkably different turbulence state ($r_{A}=0.84$) with compressible steepening, showing that the shock's efficiency at injecting compressible energy is not uniform across the entire front. Therefore, observation suggests the ICME shock front is heterogeneous even across a very small spatial scale of $\approx 100 R_{E}$ and demands multispacecraft observations (for example \cite{Trotta_2023}).\\

As mentioned earlier, turbulence inside the MC displays a Kolmogorov-scale cascade \citep{Shaikh_2024} in the inertial range. The MCs evolve since their eruption from the Sun, allowing sufficient time for the turbulence to process any initial fluctuations into the universal cascade, unless internal discontinuities and interactions among multiple ICMEs are present. The component anisotropic scaling ratio is close to one for almost all spacecraft (e.g., Wind at 0.92, DSCOVR at 0.95, ACE at 1.01), indicating developed turbulence. Except at Aditya-L1, which shows strong field-aligned anisotropy($\alpha_{\parallel}=-2.00$). This suggests that even within the coherent MC, certain sectors can show strong anisotropies. The possible reason, as we pointed out earlier, might be multiple ICME interactions or internal disruptions that intensify localized damping within the flux rope. In this context, we examined a particular region, as mentioned in section \ref{sec:results}, which showed intense turbulent power around 06:00 UT on October 11 within the MC.


\begin{deluxetable*}{lcc|cc|cc}[ht!]
\tablecaption{Spectral slopes of $B_{\parallel}$ and $B_{\perp}$ fluctuations for different regions of the ICME.\label{tab:fac_analysis}}

\tablehead{
\colhead{Satellite} &
\multicolumn{2}{c}{Solar Wind} &
\multicolumn{2}{c}{Sheath} &
\multicolumn{2}{c}{Magnetic Cloud (MC)} \\
\cline{2-7}
& \colhead{$\alpha_{\parallel}$} & \colhead{$\alpha_{\perp}$}
& \colhead{$\alpha_{\parallel}$} & \colhead{$\alpha_{\perp}$}
& \colhead{$\alpha_{\parallel}$} & \colhead{$\alpha_{\perp}$}
}

\startdata
Wind      & $-1.98 \pm 0.04$ & $-1.68 \pm 0.03$ & $-1.82 \pm 0.03$ & $-1.53 \pm 0.02$ & $-1.75 \pm 0.02$ & $-1.61 \pm 0.01$ \\
DSCOVR    & $-2.09 \pm 0.07$ & $-1.85 \pm 0.04$ & $-1.15 \pm 0.05$ & $-1.56 \pm 0.03$ & $-1.76 \pm 0.04$ & $-1.68 \pm 0.02$ \\
ACE       & $-1.66 \pm 0.05$ & $-1.68 \pm 0.03$ & $-1.44 \pm 0.04$ & $-1.72 \pm 0.03$ & $-1.66 \pm 0.05$ & $-1.68 \pm 0.03$ \\
Aditya-L1 & $-1.53 \pm 0.03$ & $-1.49 \pm 0.03$ & $-1.50 \pm 0.04$ & $-1.74 \pm 0.03$ & $-2.00 \pm 0.09$ & $-1.58 \pm 0.02$ \\
\enddata

\end{deluxetable*}

\subsection{Power spectral Analysis of an interaction region of two MCs}

\begin{figure*}[h]
    \centering
    \includegraphics[width=\textwidth]{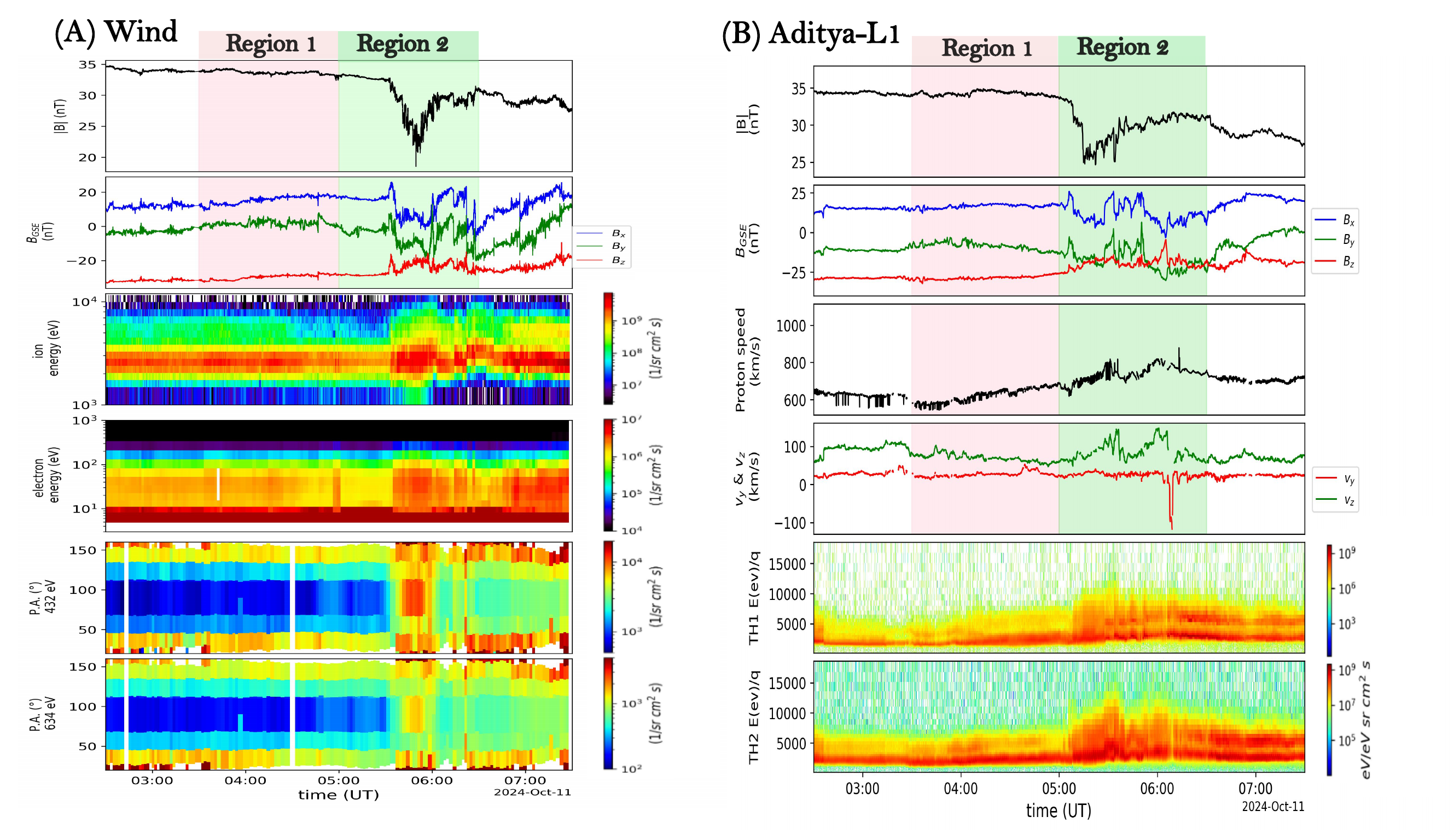}
    \caption{This figure shows magnetic field and plasma data from the two widely separated spacecrafts, Wind and Aditya-L1, in panels A and B, respectively. Both datasets demonstrate significant disruption in IMF magnitude and sudden, uneven variations in field components. The omnidirectional spectrogram of wind electrons and ions, and the suprathermal electron pitch-angle distribution at 432 and 634 eV, demonstrate particle energization and pitch-angle isotropy within MC. On a similar note, Aditya-L1 ASPEX-SWIS data show enhanced differential proton flux. Regions from 03:30:00 UT to 05:00:00 UT and from 05:00 to 6:30 UT are highlighted in pink and lime green, respectively, as regions 1 and 2.}
    \label{fig:pitchangle}
\end{figure*}

At 06:00 UT on 11th October, a sudden decrease in magnetic field intensity was observed inside the MC across all spacecraft. In addition, a localized enhancement in the bulk velocity of the solar wind, the proton temperature, and the density is evident in Figure~\ref{fig:Wind summary}. The power spectrum (Figure~\ref{fig:powerspectrum}) displays an intense magnetic fluctuation across all scales around 06:00 UT. We further investigated this region and compared the turbulence spectrum with the adjacent MC. Figure~\ref{fig:pitchangle} depicts the zoomed-in observation from the Wind spacecraft. Wind data display three successive rotations of the $B_x$ and $B_y$ components. The omnidirectional electron and ion energy flux shows a moderate increase. The pitch-angle distributions (PADs) of suprthermal electrons ($>60$ eV) generally display bidirectional field-aligned motion in the MC \citep{carcaboso2020characterisation}, which can be used as a key diagnostic for the closed magnetic topology of MC. The presence of any discontinuity within the MC makes the PAD isotropic and disrupts the coherent flux-rope structure. Moreover, we also observe field-aligned suprathermal electron flux in two energy channels, except at region 2 (Figure~\ref{fig:pitchangle}), where PAD becomes isotropic. Such evidence may indicate the presence of ongoing large-scale magnetic reconnection sites, but plasma observations remain not fully conclusive. Although observations have shown that MC internal magnetic reconnections \citep{Biswas_2025} often disrupt the coherence of the flux tube \citep{farrugia2023magnetic}. \\ 
Similar features are found in Aditya-L1 magnetometer and ASPEX-SWIS instrument data, shown in figure~\ref {fig:pitchangle}. The proton differential flux from the two top-hat electrostatic analyzers (TH1 and TH2) of ASPEX-SWIS exhibits noticeable amplification in region 2. Although our observation (the yellow-highlighted region in figure~\ref{fig:Wind summary}) and ICME catalogs \citep{https://doi.org/10.7910/dvn/c2mhth, https://doi.org/10.6084/m9.figshare.6356420} identify this event as having a single MC, this is not fully conclusive. Enhancement in the transverse component of solar wind velocity, dip in magnetic field intensity, isotropic PAD, local peaks in density, and proton-$\beta$ indicate a possible interaction region between two ejecta. However, the conclusive identification of the multiple flux ropes is out of the scope of this article.  

\begin{figure*}[h]
    \centering
    \includegraphics[width=0.9\textwidth]{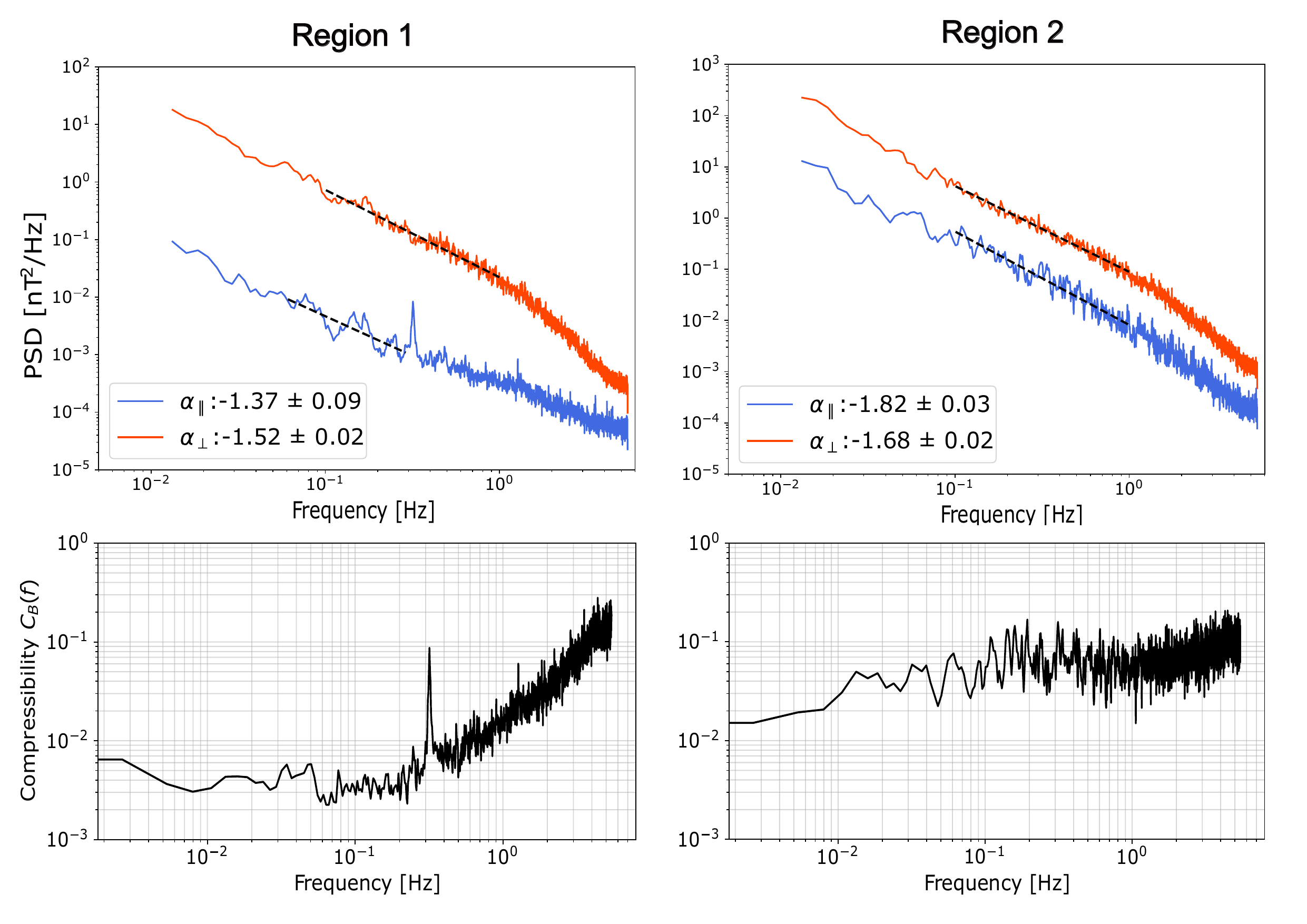}
    \caption{Zoomed-in PSD analysis highlighting Region~1 (blue) and Region~2 (red) within the magnetic cloud. Region 1 contains smooth magnetic field variation and displays significant anisotropy. Whereas region 2 shows a mature turbulence cascade. The plots below show the magnetic compressibility $C_{B}$; region 2 is highly compressible, indicating that it is dominated by intermittency and multi-scale structure.  }
    \label{fig:zoomedmcpsd}
\end{figure*}

It is well known that current sheets within MCs or interaction regions between two ICMEs lead to efficient energy injection at different scales and subsequent plasma energization through the dissipation of turbulent energy at kinetic scales. We compared the turbulent spectrum between two regions of the MC (marked by regions 1 and 2 in Figure~\ref{fig:pitchangle}). Figure~\ref{fig:zoomedmcpsd} shows the  PSD of the trace of the magnetic field obtained from Wind data. Region 1 (more structured, smoother variation) contains less turbulent power than Region 2 (discontinuity, possible current sheet), which contains higher turbulent power. Region 1, with smooth variation in the magnetic field, displays incompressible Alfvénic turbulence with IK power-law scaling in the perpendicular direction ($\alpha_{\perp}$ =- 1.52). It suggests a strong, coherent background magnetic field that inhibits the non-linear energy cascade, preventing it from reaching a fully developed state. The extremely low magnetic compressibility, $C_B(f) = {P_{|B|}(f)}/{\sum_{i=x,y,z} P_{B_i}(f)}$, confirms the dominance of incompressible Alfvenic fluctuation shown in the lower panel of figure~\ref{fig:zoomedmcpsd}. Hence, the region can be considered to be in a state of evolving turbulence. On the other hand, region 2 exhibits fully developed Kolmogorov turbulence in the perpendicular direction ($\alpha_{\perp} =- 1.68$). The parallel cascade is significantly steeper ($\alpha_{\parallel}$ =- 1.82), indicating an intermittent state due to discontinuities or an interaction region, as evident in figure~\ref{fig:pitchangle}. Furthermore, the compressibility increases by a factor of 10 relative to region 1, centered around 0.1 Hz. This strongly indicates the presence of compressible fluctuations and a region dominated by small-scale structure. These structures introduce density variations that couple with the magnetic field, a hallmark of compressible MHD. Therefore, we can precisely distinguish two distinct evolutions of the plasma state inside ICME MC.

\section{Summary and Conclusion} \label{sec:conclusion}

The azimuthal distribution of spacecrafts centered at the L1 point during the passage of ICME provided an invaluable opportunity to probe in situ signatures of plasma turbulence from different vantage points. Similar multi-point approaches have previously been shown to be essential for capturing spatial variations in ICMEs and turbulence \citep{Scolini_2024, Ghag_2024, Pal_2023, Pal_2025, lugaz2018observational}. The correlation study demonstrated that overall magnetic field measurements from different spacecraft remain highly consistent after time-lag adjustments. The near-perfect correlation in the total magnetic field strength highlights that large-scale ICME structures propagated almost coherently within approximately $100 R_{E}$. 
However, the turbulent environment of ICME substructures was distinguished by spectrogram analysis of the magnetic field. The sheath exhibited intense power, a hallmark of turbulence driven continuously by shock compression, while the magnetic cloud displayed suppressed, narrower-band power. Such contrasting behaviors are consistent with previous turbulence studies in sheath and magnetic cloud regions \citep{burlaga1991interplanetary, telloni2021spectral, Ghuge_2025}. Three regions within MC show enhanced turbulence power; one of them we investigated in detail. These results reaffirm that the sheath is the most dynamically disturbed region, whereas the MC largely preserves magnetic order, except at the discontinuities present inside the MC. A sudden, abrupt variation in the magnetic field may indicate internal reconnection or an interaction region among multiple ICMEs. \\

Spectral analysis revealed distinct differences in turbulence properties between solar wind and ICME regions. Solar wind showed the most dynamic variation across the spacecraft. We estimated the local shock impact angle ($\theta_{XN}$) using the Magnetic Coplenarity (MCo,\cite{Colburn_1966}) method. At Wind and ACE $\theta_{XN}$ (inclination of shock normal vector w.r.t GSE X direction clockwise from the Sun) are $128.4^{\circ}~and~ 135.5^{\circ}$ respectively, consistent with the interplanetary shock database \citep{Oliveira_2023}. However, at DSCOVR and Aditya-L1, the shock impact angle shows a noticeable change ($108^{\circ}$). This clearly indicates ICME shock orientation also differs in such a small segment (across $100~R_{E}$), which might play an important role in the evolution of upstream solar wind and sheath turbulence properties. In a statistical study of \cite{Borovsky_2020} showed a distinct difference between spectral slopes upstream and downstream of shocks, and in a recent study \cite{Gautam_2025} demonstrated that this shock-processing significantly alters the spectral slopes of the decoupled magnetic field components. In our multipoint observations (Figure~\ref{fig:allregionpsd}), the transition from the upstream solar wind to the downstream sheath reveals highly localized, heterogeneous shock processing, including highly variable spectral slopes in the decomposed components of magnetic field turbulence. While DSCOVR and Aditya-L1 observations strongly support these previous studies, Wind and ACE exhibit only marginal changes in the spectral slopes. We attribute this mixed behavior to the complex, non-planar geometry of the ICME shock front and to variations in its impact angle even over small separations of tens of $R_{E}$. Moreover, future multipoint analysis could benefit from applying mode decomposition techniques to separate total magnetic fluctuations into distinct magnetic field modes, as demonstrated in \cite{Gautam_2025}, to examine the evolution of spectral slopes. 

\begin{figure*}[h]
\centering
\includegraphics[width=0.8\textwidth]{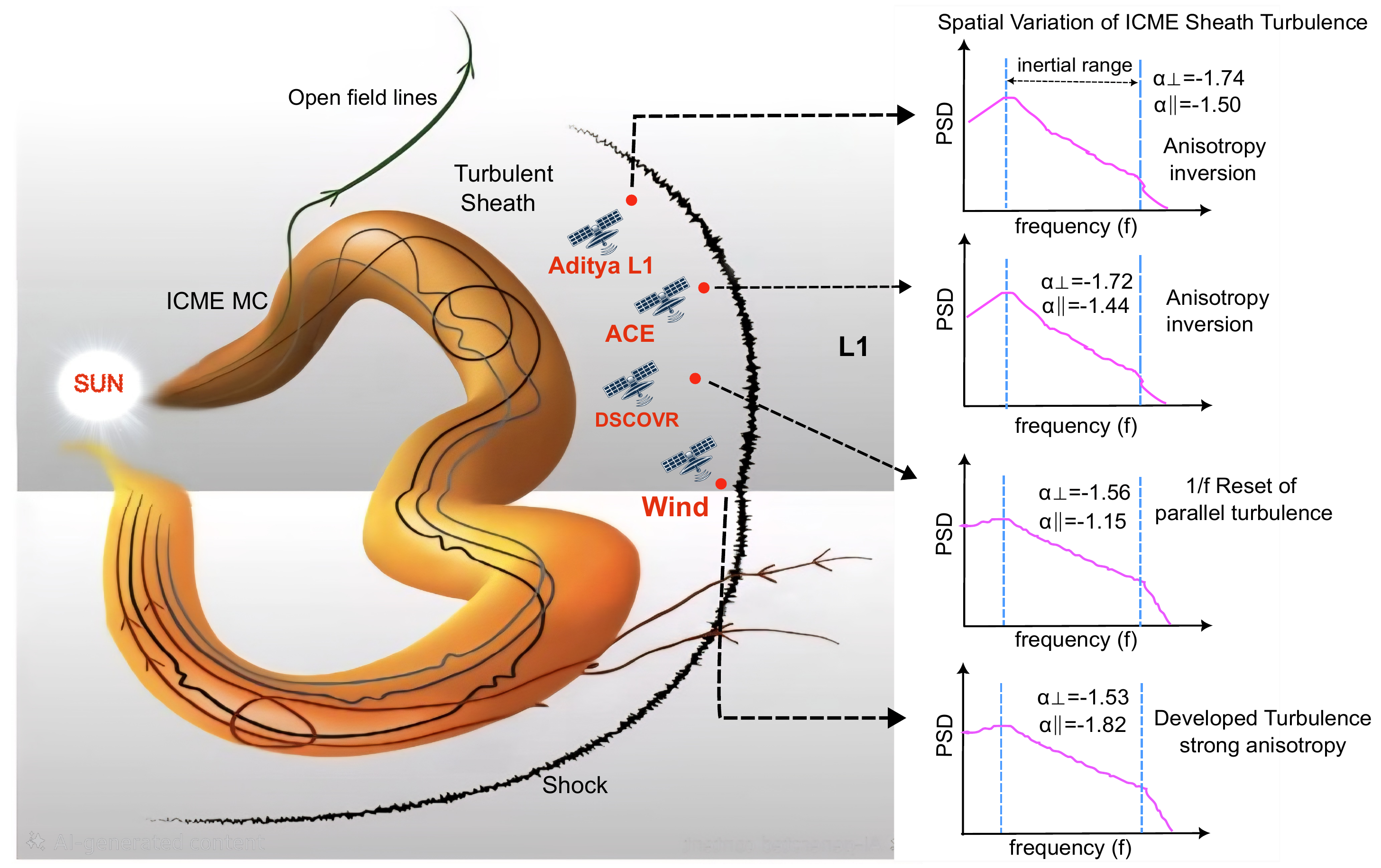}
\caption{Schematic illustration of variability of ICME turbulence anisotropy across all four spacecrafts. The ICME figure is adapted from \cite{Al_Haddad_2025} and further modified using AI. Although the picture shows a single ICME, multiple ICMEs interacted during the October 2024 extreme solar storm. The figure is not to scale.}
\label{fig:Drawing}
\end{figure*}

However, the investigation of turbulence characteristics across all regions and spacecraft in the field-aligned coordinates yielded the most interesting results. The ambient solar wind is a patchwork, showing greater variability and a more anisotropic character in the Dusk sector (-ve GSE-y). On the other hand, ACE exhibits isotropic Kolmogorov steady-state turbulence. Aditya-L1 (furthest Dawnside) magnetic turbulence is dynamically aligned with Boldyrev or IK power-law scaling \citep{Boldyrev_2006, iroshnikov1964turbulence}; therefore, the two components are locked into the same evolutionary scaling. The ICME sheath is characterized by fundamental asymmetry. The shock preferentially injects energy into the compressible modes, resulting in a \enquote{young}, shallow $\alpha_{\parallel}$ spectrum, while the incompressible modes continue to cascade efficiently. Three out of four spacecraft (ACE, DSCOVR, and Aditya-L1) exhibit anisotropic inversion, suggesting sheath turbulence is strongly driven by shock compression. The parallel-direction turbulence is reset to nearly $f^{-1}$ state. Figure~\ref{fig:Drawing} illustrates the variation of sheath turbulence around the L1 point. Finally, MC acts as a turbulent homogenizer, forcing the chaotic, decoupled compressible and incompressible fluctuations of the sheath into a unified, mature state where both component cascades occur at similar Kolmogorov-like rates.  

Eventually, as we focus on the local scale within MC, we observe a departure from the steady-state Kolmogorov nature and a shallower spectral slope (region 1). Further investigation reveals the region is not Alfv\'enic (not shown). A recent statistical survey of \cite{Kumbhar_2026} also found that more than half of MC regions exhibit non-Alfvenicity, which may lead to contrasting turbulence properties. On the other hand, region 2 is highly turbulent. The abrupt variation in the magnetic field and the energization of electrons and ions, with an isotropic suprathermal electron pitch-angle distribution, reveal reconnection signatures (interaction region between multiple MCs). These regions are of great interest to turbulence studies because of their multiscale structures, which lead to particle acceleration and heating through wave-particle interactions. The magnetic compressibility is higher by an order of magnitude, confirming the dominance of compressible turbulence.\\

In conclusion, this study presents the first observational evidence of varying turbulence states within the ICME sheath and MC, even across very small azimuthal angular separations, using multipoint L1 solar wind observations, including measurements from ISRO's Aditya-L1. The results provide strong evidence of intrinsic heterogeneity in plasma conditions across different regions of an ICME downstream of the shock. Therefore, these regions should be considered separately in future studies aimed at determining the true downstream spectral behavior, which is critical to improving the reliability of space weather modeling and forecasting. The solar wind–magnetosphere coupling at the dayside magnetopause is highly sensitive to the turbulent solar wind environment \citep{Gu_2025}. In this context, pronounced spatial inhomogeneity within the ICME sheath and MC can potentially enhance or suppress energy transfer into the magnetosphere by modulating the efficiency of magnetic reconnection. Consequently, multipoint observations of ICME turbulence and its properties near the Sun-Earth L1 point are essential for accurately assessing their geoeffectiveness and improving predictive space weather frameworks. The present space era offers a golden opportunity for multi-point in situ observations of the solar wind near L1, as several heliophysics missions from ISRO, NASA, and NOAA are actively observing it. Future coordinated observations by this constellation of multiple spacecraft around the Sun-Earth L1 point will enhance our understanding of the solar wind near Earth and improve our ability to forecast space weather.


\section{Accknowledgement} \label{sec:accknowledgement}
The authors acknowledge the ACE, Wind, DSCOVR, and Aditya-L1 teams for providing scientific data for this study. S.B. acknowledges the research grant provided by the Indian Space Research Organization as a research fellowship. We sincerely acknowledge the support of Dr. Simon Good for his assistance with the interpretation of turbulence anisotropy. The scientific figures have been made using the pySPEDAS (\url{https://github.com/spedas/pyspedas}) module of Python. The authors also acknowledge the Community Coordinated Modeling Center (CCMC) for the simulation (\url{https://ccmc.gsfc.nasa.gov/}).   

\bibliographystyle{aasjournal}
\bibliography{bibloigraphy}

\end{document}